\documentclass[aps,prc,twocolumn,showpacs,floatfix,nofootinbib,preprintnumbers,superscriptaddress,amsmath,amssymb,longbibliography,amsproc]{revtex4-1}
\usepackage{graphicx}
\usepackage{float}
\usepackage{dcolumn}
\usepackage{multirow} 
\usepackage{hyperref}
\usepackage{nicefrac}
\graphicspath{{.}{./../figures/}{./../notebooks/}}

\newcommand{\be}{\begin{equation}}
\newcommand{\ee}{\end{equation}}
\newcommand{\ba}{\begin{eqnarray}}
\newcommand{\ea}{\end{eqnarray}}

\begin{document}

\title{Extrapolation of nuclear structure observables with artificial
  neural networks}

\author{W.G. Jiang}
\affiliation{Department of Physics and
   Astronomy, University of Tennessee, Knoxville, TN 37996, USA}
\affiliation{Physics Division, Oak Ridge National Laboratory, Oak
   Ridge, TN 37831, USA}

\author{G.~Hagen}
\affiliation{Physics Division, Oak Ridge National
  Laboratory, Oak Ridge, TN 37831, USA}
\affiliation{Department of
  Physics and Astronomy, University of Tennessee, Knoxville, TN
   37996, USA}
   
\author{T.~Papenbrock}
\affiliation{Department of Physics and
   Astronomy, University of Tennessee, Knoxville, TN 37996, USA}
\affiliation{Physics Division, Oak Ridge National Laboratory, Oak
   Ridge, TN 37831, USA}

\begin{abstract}
  Calculations of nuclei are often carried out in finite model
  spaces. Thus, finite-size corrections enter, and it is necessary to
  extrapolate the computed observables to infinite model spaces.  In
  this work, we employ extrapolation methods based on artificial
  neural networks for observables such as the ground-state energy and
  the point-proton radius.  We extrapolate results from no-core shell
  model (NCSM) and coupled-cluster (CC) calculations to very large
  model spaces and estimate uncertainties. Training the network on
  different data typically yields extrapolation results that cluster
  around distinct values.  We show that a preprocessing of input data,
  and the inclusion of correlations among the input data, reduces the
  problem of multiple solutions and yields more stable extrapolated
  results and consistent uncertainty estimates.  We perform
  extrapolations for ground-state energies and radii in $^{4}$He,
  $^{6}$Li, and $^{16}$O, and compare the predictions from neural
  networks with results from infrared extrapolations.
\end{abstract}


\maketitle 

\section{Introduction}
In nuclear physics, {\it ab initio} methods aim to solve the nuclear
many-body problem starting from Hamiltonians with two- and
three-nucleon forces using controlled
approximations~\cite{dickhoff2004,navratil2009,lee2009,barrett2013,roth2012,dytrych2013,hagen2014,carlson2015,hergert2016}. Most
of these methods employ finite model spaces, and this makes it
necessary to account for finite-size corrections or to extrapolate the
results to infinite model spaces. While light nuclei with large
separation energies require little or no extrapolations, finite-size
effects are non-negligible in weakly bound nuclei or heavy nuclei.
Various empirical extrapolation schemes
\cite{horoi1999,zhan2004,hagen2007b,forssen2008,bogner2008} have been
used. More recently, rigorous extrapolation formulas were derived
based on an understanding of the infrared and ultraviolet cutoffs of
the harmonic oscillator
basis~\cite{furnstahl2012,coon2012,more2013,furnstahl2014,konig2014,wendt2015,odell2016,forssen2018}.
These extrapolation formulas are akin to L{\"u}schers
formula~\cite{luscher1985} derived for the lattice and its
extension~\cite{konig2017} to many-body systems.  Unlike the lattice,
however, the harmonic oscillator basis mixes ultraviolet and infrared
cutoffs, and this complicates extrapolations.  Very recently, Negoita
and coworkers~\cite{negoita2018a,negoita2018} employed artificial
neural networks for extrapolations. They trained a network on NCSM
results obtained in various model spaces, i.e. for various oscillator
spacings $\hbar\omega$ and different numbers $N_{\rm max}\hbar\omega$
of maximum excitation energies. In practical calculations, $N_{\rm
  max}\approx 10 \ldots 20$ in light nuclei. The neural network then
predicted extrapolations in very large model spaces of size $N_{\rm
  max}\sim 100$. Impressively, the neural network also predicted that
the ground-state energies and radii cease to depend on the oscillator
spacing as $N_{\rm max}$ increases. Negoita and coworkers employed
about 100 neural networks, each differed by the initial set of
parameters (weights) from which the training started. The resulting
distributions for observables occasionally exhibited a multi-mode
structure stemming from multiple distinct solutions the neural
networks arrived at. In this work, we want to address this challenge
and focus on the network robustness and avoidance of multiple
solutions.

In recent years, artificial neural networks have been used for various
extrapolations in nuclear
physics~\cite{clark2001,athanassopoulos2004,costiris2009,akkoyun2013,utama2016,utama2016b,utama2017,utama2018,neufcourt2018},
and for the solution of the quantum many-body
system~\cite{carleo2017}.  Artificial neural networks use sets of
nonlinear functions to describe the complex relationships between
input and output variables. The universality of using artificial
neural networks to solve extrapolation problems is largely guaranteed,
because no particular analytical functions are needed. Artificial
neural networks are controlled by two hyperparameters, i.e. the number
of layers and the number of neurons for each layer.

There are still two major challenges when introducing neural networks
in extrapolations of results from {\it ab initio}
computations. Firstly, unlike other applications in which large
amounts of training data can be acquired, the inputs provided by the
{\it ab initio} calculations are limited to small data sets. The
statistics is clearly not enough to support the network training
without overfitting. Secondly, randomness, caused by the nature of
basic network algorithms, is an intrinsic quality of the neural
network that conflicts with the high-precision requirement for
extrapolations.

In this work, we use an artificial neural network and extrapolate
observables computed with the NCSM and CC methods. Besides standard
techniques such as regularization, we use interpolation of data to
mitigate the overfitting problem and also take into account the
correlations in the resulting data set.  The random initialization of
the network parameters provides us with a ``forest'' of
artificial neural networks.  This allows us to gain insights
into uncertainties of the extrapolated observables, 
under the precondition that the distribution of extrapolation results
has a single peak.

We note here that the extrapolation problem we are concerned with is
special in the sense that a well defined asymptotic value exists for
the observable of interest (i.e. an energy or a radius), that there is
a simple pattern in the learning data, and that the learning data is
already close to this asymptotic value. We will see below that this
makes an artificial neural network a useful tool for this kind of
extrapolation. Needless to say, for a general problem there is no tool
to extrapolate: we cannot extrapolate from available data to next
week's stock market value or next month's weather. We refer the reader
to the literature for attempts to use deep learning in
extrapolations~\cite{martius2016}, and for a counter
example~\cite{haley1992}.

This paper is organized as follows. In the next Section we introduce
the theoretical framework and artificial neural networks and present a
detailed account of how we construct, train, and use neural
networks. We then present and discuss the extrapolation results for
$^4$He, $^6$Li, and $^{16}$O. Finally, we summarize our work.

\section{Theoretical Framework}

\subsection{Artificial Neural Network Architecture}
An artificial neural network is a computing system that consists of a
number of interconnected blocks which process the input information
and yield an output signal.  Modeled loosely after the human brain,
the neural network is typically organized by similar blocks called
``layers,'' and each layer contains a certain number of parallel
``neurons.'' The numbers of layers and neurons define the depth and
the width of the neural network, respectively.

\begin{figure}
  \includegraphics[width=0.47\textwidth]{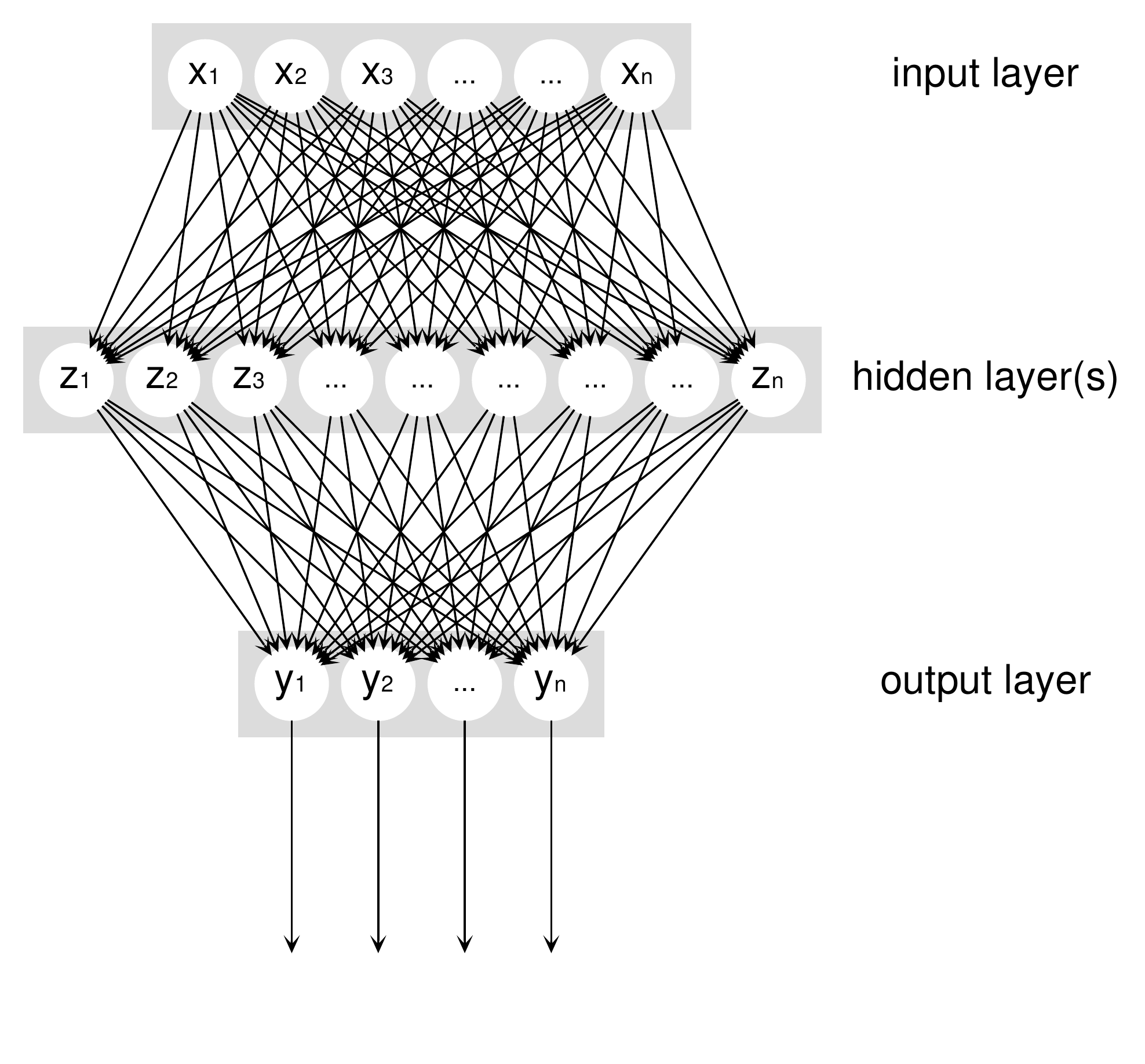}
  \caption{(Color online) Schematic structure of a typical feed-forward neural network. }
  \label{fig:NN_schematic_diagram}
\end{figure}

Figure~\ref{fig:NN_schematic_diagram} shows the schematic structure of a
simple feed-forward neural network. The algorithm basically consists of two
parts. First, the input signal $x$ is propagated to the output layer
$y$ by a series of transformations. The whole network can be seen as a
complex function between the input and output variables. In the simple case
with one hidden layer, the function can be written as follow,
\begin{eqnarray}
z_{j} = \displaystyle\sum_{i}x_{i} w_{ij}+b_{j},
\end{eqnarray}
with $\sigma$ as the activation function,
\begin{eqnarray}
x'_{j}= \sigma(z_{j})
\end{eqnarray}
\begin{eqnarray}
y_{k}= \displaystyle\sum_{j}x'_{j} w'_{jk}+b'_{k}.
\end{eqnarray}
Here, $x_{i}$ are the input variables, and $y_{k}$ are the output
variables. The weights $w~(w')$ and bias $b~(b')$ are free parameters
of the neural network. There exist a few choices one can make for the
activation function $\sigma$, such as the sigmoid, tanh and
Rectified~Linear~units (ReLu). These are non-linear functions which
enable the neural network to capture complex non-linear relationships
between variables. For the extrapolation we follow
Ref.~\cite{negoita2018} and use a smooth activation function that only
acts on the hidden layer.
Back-propagation is the second part of the
algorithm~\cite{rumelhart1986}. This is the central mechanism that
allows neural network methods to ``learn.'' The error signals, often
referred to as the ``loss,'' which measure the deviation between the
predicted output $y_{\rm{pre}}$ and the training target
$y_{\rm{true}}$, are propagated backwards to all the parameters of the
network and allow the optimizer to update the network status
accordingly. Note that, in practice, the neural network always
processes the data in batches, which makes the input (output) signals
$x~(y)$ matrices and the network functions become matrix operations.

In order to construct the artificial neural network aiming to solve
the extrapolation problem, we first need to determine its topological
structure. There are a lot of variants for neural networks, such as
Recurrent Neural Network (RNN), Long Short-Term Memory (LSTM) and
Convolutional Neural Network (CNN), which are designed for various
assignments. One should choose the appropriate type of network
according to the organizational structure of the dataset and the goal
that one wants to achieve. In the case of extrapolation, the data for
training is assigned to a structure consisting of three members,
namely $\hbar \omega$, $N_{\rm{max}}$, and the corresponding target
observables, i.e. the ground-state energy and the point-proton
radius. On the other hand, the main purpose of the neural network is
to provide reasonable predictions for the observables at any values of
$\hbar \omega$ and $N_{\rm{max}}$. In this paper we use the
feed-forward neural network, which takes the $\hbar \omega$ and
$N_{\rm{max}}$ as two inputs $(x)$ and the target observables as
output $(y_{\rm{true}})$.  One could as well apply the RNN structure
to achieve the same goal. The only difference between the two choices
is that the data structure need to be reorganized in terms of
sequential observable values with increasing $N_{\rm{max}}$ under the
same $\hbar \omega$.

Once the basic structure is decided, the next task is to control the
complexity of the network. The network's ability of describing complex
features is determined by the numbers of the hidden layers and neurons
in each layer. In other words, the depth and the width of the neural
network control the upper limit of the neural network
description. Ideally, in order to lower the loss of the training
dataset, adding more layers and neurons is always helpful to incease
its accuracy. However, as the neural network becomes more complex it
becomes harder to train. Given the same amount of training data, a
deeper and wider network requires more time to get converged results,
and one risks overfitting of the network's parameters. In extreme
cases, for instance, when the network is so complex that it has much
more parameters than the number of input data, it can easily get 100\%
of accuracy on the training set, but still perform poorly on testing
samples. Instead of learning the pattern, the network simply memorizes
the training data and exhibits no predictive power.

Even though there is no exact answer for how to configure the numbers
of layers and neurons in the neural network, there are still some
guiding principles to follow. For a start, we consider a network with
one hidden layer.  Based on the universal approximation
theorem~\cite{cybenko1989,funahashi1989,hornik1991} any continuous
function can be realized by a network with one hidden layer.  Of
course, a deep neural network (with multiple hidden layers) will have
certain advantages over the shallow one (with few hidden layers). For
example, the deep neural network can reach the same accuracy of a
shallow one with much fewer
parameters~\cite{bengio2009,seide2011,mhaskar2016}. However, in order
to prevent problems such as vanishing gradients and overfitting, the
architecture of the deep neural network needs careful construction
including, but not limited to: initialization of the network
parameters~\cite{glorot2010}, design of the activation
function~\cite{glorot2011}, using the proper
optimizer~\cite{kingma2014}, and improving the training
procedure~\cite{bengio2007}. For our task of extrapolation, a deep
neural network would be an overkill. As for the numbers of neurons,
there are several empirical rules~\cite{stathakis2009} and techniques,
such as pruning~\cite{heaton2008} that can be applied. In the present
work, we start with a simple structure and then increase the numbers
of neurons and layers until we arrive at a sufficiently small loss for
the training dataset.  For the results shown below, we arrived at
neural networks with a single hidden layer, consisting of eight and 16
nodes for the extrapolation of energies and radii, respectively.

\begin{figure}[b]
  \includegraphics[width=0.43\textwidth]{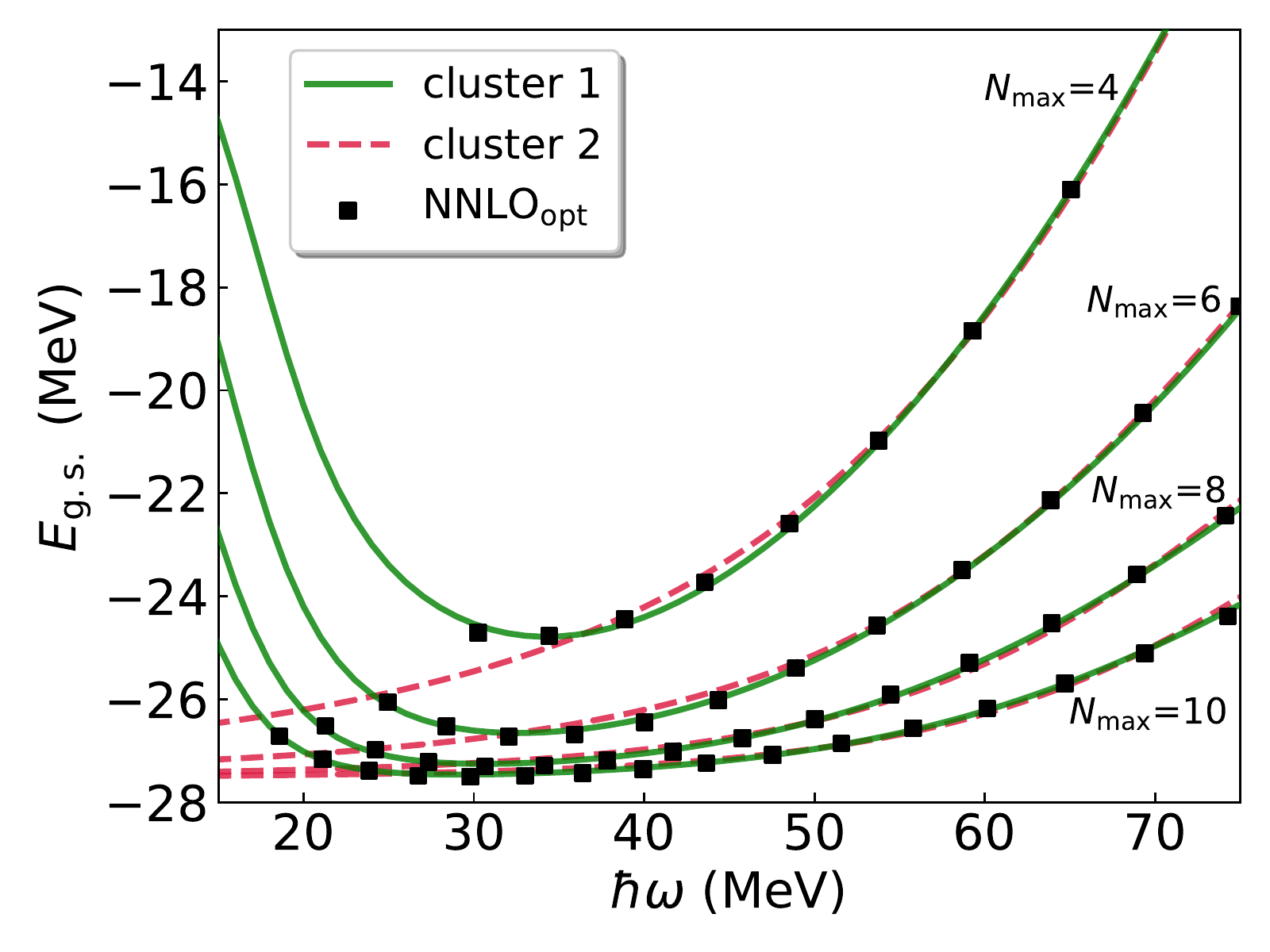}
  \caption{(Color online) Ground-state energies from NCSM computations
    of $^4$He based on the NNLO$_{\rm opt}$ potential (black data
    points). The green full line and the red dashed line show two
    different neural network solutions for learning the ground-state
    energy of $^4$He in finite model spaces. }
  \label{fig:cluster_compare}
\end{figure}

Figure ~\ref{fig:cluster_compare} shows some of the data we used in
extrapolations of the ground-state energy of $^4$He. The black points,
taken from Ref.~\cite{forssen2018}, denote results from NCSM
computations based on the NNLO$_{\rm opt}$
potential~\cite{ekstrom2013}. The ground-state energies are shown as a
function of the oscillator frequency and labeled by the number $N_{\rm
  max}$ of employed oscillator excitations.

Figure ~\ref{fig:cluster_compare} also shows that the data exhibits a simple pattern, namely U-shaped
curves that get wider and move closer together as $N_{\rm max}$
increases (See also Fig.~\ref{fig:different_Nmax_observables_O16} for
another example.) To capture this behavior with an artificial neural
network, we choose a sigmoid as the activation function,
i.e. $\sigma(x)=(1+e^{-x})^{-1}$. It is then clear that asymptotic
values of large $N_{\rm max}$ either map to zero or to one in the
activation function, and this explains why -- by design -- an
asymptotically flat function results in the extrapolation. Indeed,
using a ReLu function as the activation function
(i.e. $\sigma(x)=\max{(0,x)}$) leads to noisy extrapolation results.

\subsection{Data Interpolation and Correlated Loss}
Despite the fact that we can easily design a neural network that gives
satisfactory accuracy on training data, a good performance on making
predictions is not guaranteed for the extrapolation problem. More
often than not the loss of the testing data will be much larger than
the loss of the training data, which is a clear sign of
overfitting. Overfitting is a major issue for neural network
applications, which is usually caused by the conflict between having
insufficient information from a limited dataset, and the networks
flexibility to approximate complex non-linear functions. This is
exactly the case for the {\it ab initio} extrapolation task at hand.
The {\it ab initio} calculations are restricted to a not-too-large
value of $N_{\rm{max}}$, and for a given $N_{\rm max}$ only a few
oscillator spacings $\hbar\omega$ are available.  In the case of
$^4$He, for instance, we only have 144 data points from NCSM
calculations, and this is is inadequate for training even a very
simple neural network, thus overfitting seems inevitable.

There are a few strategies that can be introduced to avoid overfitting
in neural networks, including regularizations~\cite{zou2005},
dropout~\cite{srivastava2014}, and early
stopping~\cite{prechelt1998}. Such methods can be used together or
separately to increase the network robustness and reduce
generalization errors. The price to pay is that one will have to deal
with more hyperparameters and determine the best combination of
them. Besides these methods, one of the best ways to reduce
overfitting is to enlarge the data set. In our case, however, the
commonly used practice of data augmentation~\cite{tanner1987} and
addition of random noise to the data set will not be helpful, because
extrapolation is a quantitative problem that requires high accuracy
and input data with a clear physical foundation.

To enlarge the data set, we note that the {\it ab initio} calculations
for a given $N_{\rm{max}}$ should give a continuous smooth curve for
the target observable values as a function of $\hbar\omega$. The
limited input data is merely restricted by the computation cost but
not by the method itself. Thus, performing interpolation on existing
data is an economical way to obtain more information.  In this work,
we employ a quadratic spline for interpolation in $\hbar\omega$ at
fixed $N_{\rm max}$.  This procedure increases the robustness of the
neural network even with the basic single-hidden-layer architecture
and avoids overfitting.

As a large portion of the training data is generated by interpolation,
the standard ``$\chi^2$'' loss function (valid for independent data)
might not be appropriate. As the generation of $n$ points via
interpolation yields $n$ correlated samples, we introduce the
correlated loss function
\begin{eqnarray}
  \label{losscorr}
  L = \displaystyle\sum_{i=1}^{n}\displaystyle\sum_{j=1}^{n} W_{ij}R_{i}R_{j} .
\end{eqnarray}
Here $W_{ij}$ are the elements of a correlation matrix, and $R_{i}$
$(R_{j})$ are the residuals of the $y_{\rm{pre}}$ and the target
$y_{\rm{true}}$. In this work, we will either consider the absence of
correlations (i.e. $W_{ij}=\delta_{ij}$) or include correlations as
described in what follows. The elements $W_{ij}$ form a block matrix,
because only points interpolated at fixed $N_{\rm max}$ are correlated
by the spline.  For fixed $N_{\rm{max}}$ the block matrix is taken to
be tridiagonal with all non-zero matrix elements equal to one. This
indicates that the correlation is only between neighboring data
points. We note that the loss function~(\ref{losscorr}) is usually not
a built-in function for much of the mainstream neural network
development environments. Thus, we employ a customized loss function,
and the position $i$ or $j$ of each data point is needed as an
additional input for the network to generate the correlation matrix
with elements $W_{ij}$.


Training a neural network starts with a random initialization of the
network parameters (weights and biases). During training the loss
function is minimized using the training data set as input. It is
clear that the random starting points will lead to different trained
networks, because optimizers can generally not find the global minimum
of the loss function. The existence of many local minima with an
acceptable loss will thus lead to different network predictions.

Inspired by the random forest algorithm \cite{breiman2001}, in which
the decision forest always gives better performance than a single
decision tree, we introduce multiple neural networks with the same
structure but with different initialized parameters to address the
uncertainty problem. The outputs of all the networks are being
integrated in order to obtain a range of predictions and uncertainty
estimates.  This approach is going to help us to reveal some insights
into neural networks, and guide us in selecting favorable neural
network solution.

\begin{figure}[htb]
  \includegraphics[width=0.52\textwidth]{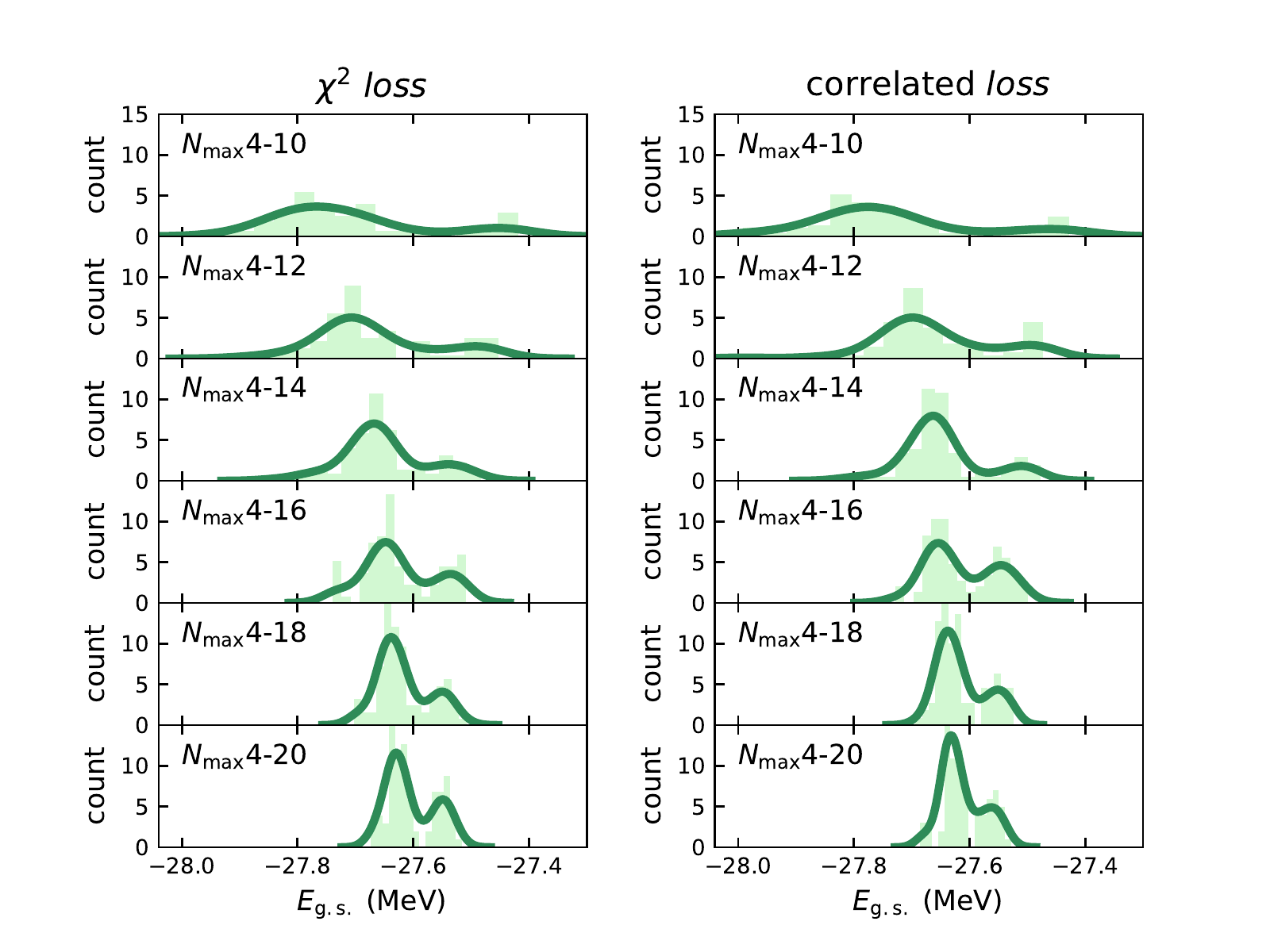}
  \caption{(Color online) Distributions of multiple neural network
    trained with different max$(N_{\rm{max}})$ datasets for
    ground-state energy of $^4$He using $\chi^2$ loss function (left
    panel) and correlated loss (right panel). }
  \label{fig:multi-NN_distribution}
\end{figure}

Figure~\ref{fig:multi-NN_distribution} demonstrates the impact of
including correlations into the loss function. The left panels shows
the predictions of 100 neural networks for the ground-state energy of
$^4$He. The input data consists of NCSM data for model spaces with a
maximum value of $N_{\rm max}$ as indicated, and the correlation
matrix $W$ of Eq.~(\ref{losscorr}) is taken to be diagonal, i.e. no
correlations are included. The displayed ground-state energies are the
neural network predictions for $N_{\rm max}=100$, and there is
virtually no $hw$-dependence.  The shown distribution function results
from Kernel Density Estimations (KDE), i.e. by replacing the
delta-function corresponding to each individual data point with a
Gaussian. The distribution becomes narrower as the input data includes
increasing values of $N_{\rm max}$. We note that the distributions are
bi-modal.

The inclusion of correlations, shown in the right panel of
Fig.~\ref{fig:multi-NN_distribution}, somewhat reduces the importance
of the smaller peak. The main peaks, which include most of the network
results, exhibit a smaller average loss and therefore are believed to
be the better solution. Their central values are likely the to be the best
predictions for these networks. However, for uncorrelated and
correlated loss functions, the second peak does not appear by accident
and can not be neglected. Its persistence against different optimizers
and hyperparameter adjustments shows that it is a stable local minimum
and not too narrow. From this point of view, both peaks can be treated
as the solutions of the multiple neural networks. As the maximum
$N_{\rm{max}}$ of the input data is increased, the two peaks are
getting closer to each other but remain distinguishable. Thus, a
significant uncertainty remains.

\subsection{Multiple Neural Network and Data Preprocessing}

\begin{figure}
  \includegraphics[width=0.45\textwidth]{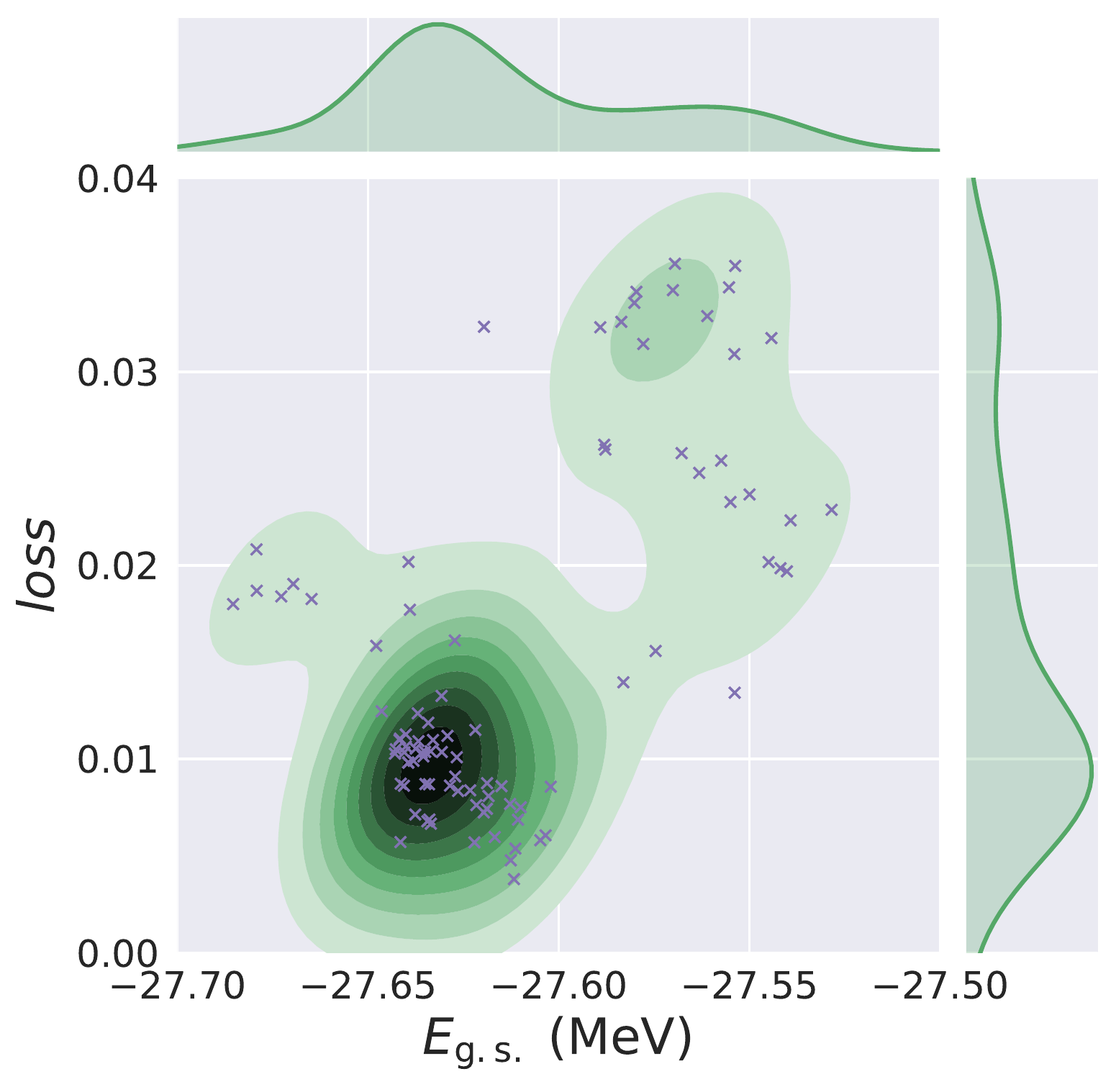}
  \caption{(Color online) Multiple neural network predictions for extrapolating
    ground-state energy of $^4$He with NCSM calculated dataset
    max$(N_{\rm{max}})=20$ as input. Kernel density estimations for
    $E_{\rm{g.s.}}$ and loss are also given along side. The
    calculation contains 100 independent random initialized neural network. }
  \label{fig:multi_NN}
\end{figure}

We want to understand the bi-modal structure of the distribution
functions. For this purpose, we focus on the correlated loss function.
Figure~\ref{fig:multi_NN} presents results from 100 neural networks
for the correlated loss versus the ${^4}$He ground-state energy
$E_{\rm{g.s.}}$.  Each cross in Fig.~\ref{fig:multi_NN} represents one
fully trained neural network and has already reached convergence
(i.e. the loss shift is within a required accuracy). As before, the
shown distribution function results from KDE. Each individual data
point (crosses) and contour lines are also shown.  The top and right
panels show the integrated distributions for the ground-state energy
and the loss, respectively.

We understand the double-peak structure as follows. The cluster of
networks under the dominant peak predict a U-shape for the curves
$E_{\rm g.s.}(\hbar\omega,N_{\rm max})$ at fixed $N_{\rm
  max}$. However they deviate in ``higher-order'' terms that define
the precise shape. The smaller cluster of networks under the small
peak predict curves $E_{\rm g.s.}(\hbar\omega,N_{\rm max})$ that
increase monotonically as a function of $\hbar\omega$. They have a
higher loss. This interpretation is based on the results shown in
Fig.~\ref{fig:cluster_compare}. Here, the black squares are the input
data of ground-state energies for given $\hbar\omega$ and $N_{\rm
  max}$. The green full lines show predictions from the first cluster
of networks under the dominant peak of Fig.~\ref{fig:multi_NN}. In
contrast, the red dashed lines are predictions from the second cluster
of networks under the smaller peak in Fig.~\ref{fig:multi_NN}.  It is
evident that the networks of cluster 1 learned the pattern of all data
while those of cluster 2 failed to predict the trend of the data
points at smaller $\hbar\omega$. How did the neural networks of
cluster 2 make this mistake?

Inspection showed that the imbalanced dataset is the root of the
problem. Our dataset includes many points at relatively large
$\hbar\omega$ values (as we used such ultraviolet converged points for
infrared extrapolations in Ref.~\cite{forssen2018}), and the
corresponding ground-state energies are also much above the
variational minimum and the infinite-space result. In contrast, the
data set contains a smaller number of data points at relatively small
values of $\hbar\omega$, and the corresponding ground-state energies
are much closer to the infinite-space result. Thus, the failure to
correctly learn about these ``minority'' data points yields a
relatively small increase of the loss function.  With random
parameters initialization, once the network reaches a local minimum,
the imbalanced dataset will, to a large extent, prevent the optimizer
from pulling the network out of it. Furthermore, with the imbalanced
training data, the effort of emphasizing the minority data directly
conflicts with the idea of reducing overfitting. Some of the common
neural network strategies, such as adding a regularization term, will
make things worse. In contrast, removing data points at too large
values of $\hbar\omega$ from the training data set, or a stronger
weighting of data closer to the variational minimum (at fixed $N_{\rm
  max}$) in the loss function, reduces the number of trained networks
that would fall into cluster 2.

\begin{figure}
  \includegraphics[width=0.52\textwidth]{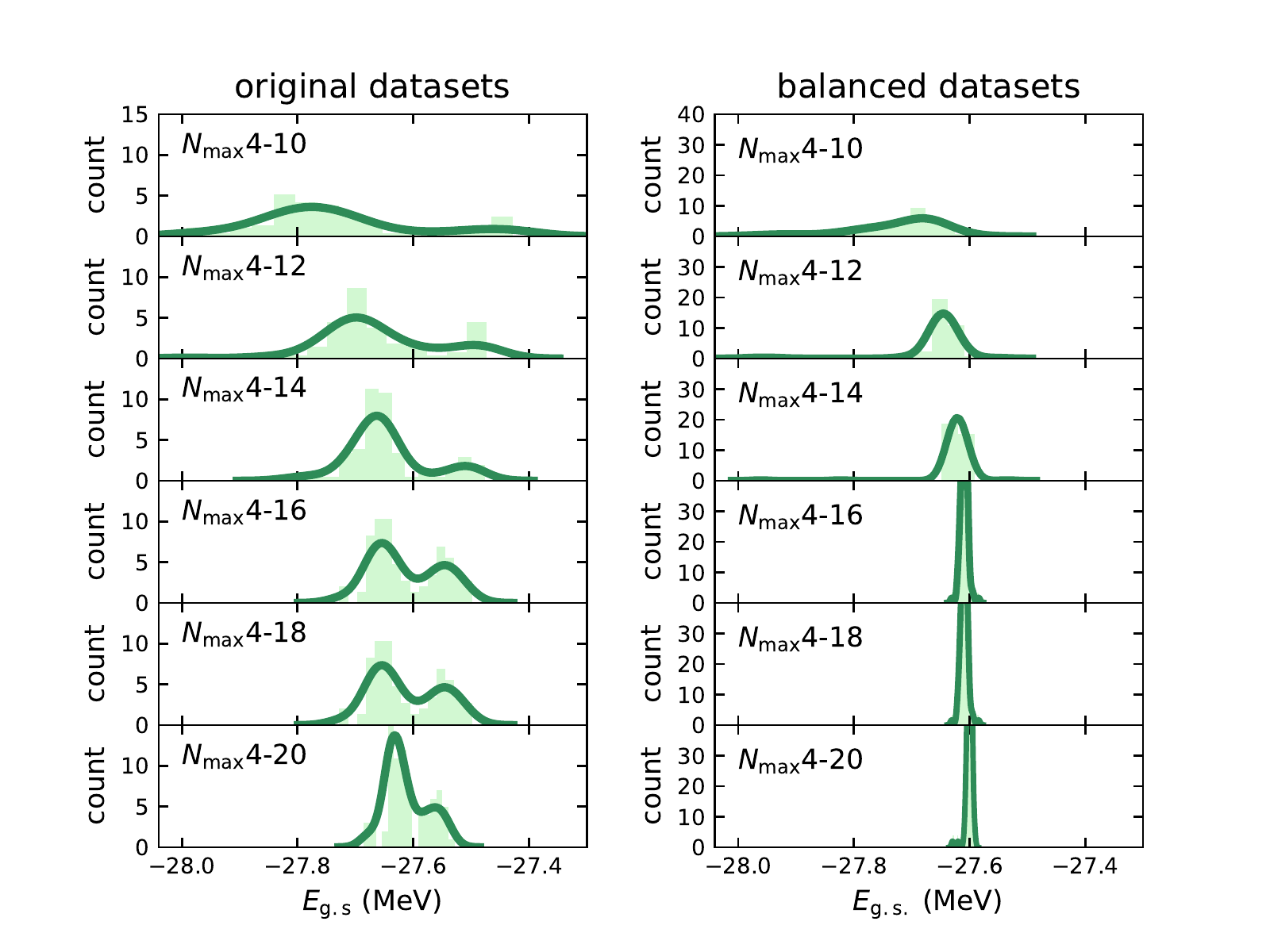}
  \caption{(Color online) Distributions of multiple neural network for
    ground-state energy of $^4$He, with the origin datasets (left
    panel) and with the preprocessed datasets (right panel).}
  \label{fig:multi-NN_distribution2}
\end{figure}

In the {\it ab initio} calculation, when the $\hbar \omega$ of the
harmonic oscillator basis is too large or too small (i.e. it deviates
from the ``optimal'' value $\hbar\omega\approx {\hbar^2\Lambda/(mR)}$,
where $\Lambda$ and $R$ are the scales set by the cutoff of the
potential and the radius of the computed nucleus~\cite{hagen2010b}),
the convergence with respect to the increasing $N_{\rm{max}}$ is slow,
because the employed basis is not efficient to capture ultraviolet and
infrared aspects of the problem.  The data points that we are most
interested in are close to the variational minimum at fixed
$N_{\rm{max}}$. To overcome the problem of the imbalanced dataset, we
apply Gaussian weights on the input data, using the values of the
minima for the centroids and a standard deviation of about 8.5~MeV.
The networks are trained using these weights and a correlated loss
function.  Figure~\ref{fig:multi-NN_distribution2} shows the
comparison of multiple neural network results with and without sample
weights. We note that the two panels have different ranges for
$y$-axis to better display the distribution of the ground-state
energy.  Training with the original datasets (left panel) yields the
bi-modal distribution.  Introducing balanced datasets via Gaussian
weights (right panel) suppresses the second peak and leaves us with
one solution for the extrapolation problem. At the same time, this
improves the precision of the predicted observable and thus yields a
smaller uncertainty for the neural network extrapolation.

\begin{figure}
  \includegraphics[width=0.48\textwidth]{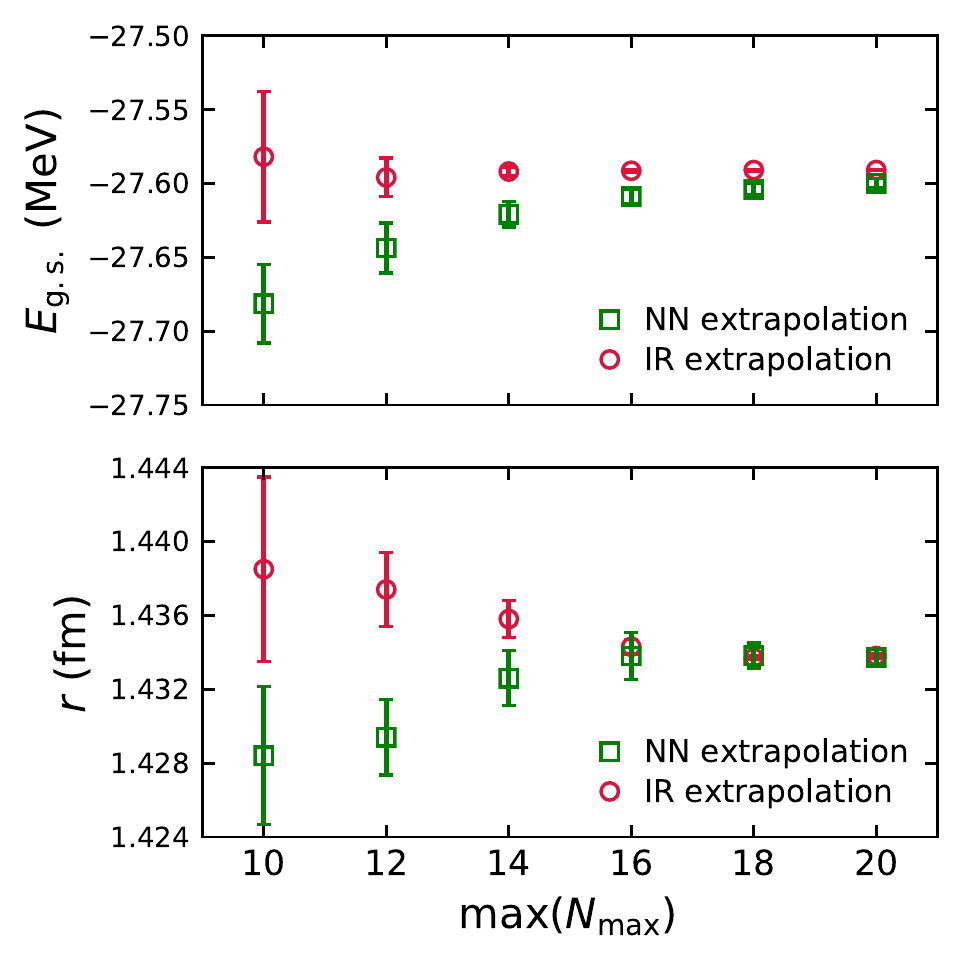}
  \caption{(Color online) Extrapolated results for $^4$He ground-state
    energy (upper panel) and point-proton radius (lower panel) with
    NCSM datasets from $\rm{max}(N_{\rm{max}})=10$ to
    $\rm{max}(N_{\rm{max}})=20$ employing neural network (squares) and
    IR (circles) extrapolation. Error bars represent the uncertainties
    of the extrapolations that are due to changes in the initial point
    in the training process.}
  \label{fig:different_Nmax_observables_He4}
\end{figure}

We note here that the increased weighting of points close to the
variational minima is a akin to employing a prior in Bayesian
statistics. Such techniques could also be used for a quantification of
uncertainities~\cite{schindler2009,furnstahl2014c,carlsson2016,neufcourt2018}. In
this work, we limit ourselves to uncertainty estimates.

\section{Results and discussions}

We now present the results of the neural networks' predictions for
ground-state energies and radii, and compare with other extrapolation
methods. We start with the nucleus $^4$He. The networks are trained
separately for the ground-state energy and radius. The datasets are
generated by NCSM calculations using the NNLO$_{\rm{opt}}$
nucleon-nucleon interaction. Since the four-nucleon bound state of
$^4$He is already well converged with the maximum model space that
NCSM calculation can reach, it is a good case to perform a benchmark
and study the performance of the neural network extrapolations. The
networks are trained with different datasets which contain the NCSM
results from $N_{\rm{max}}=4$ to the given
$\rm{max}(N_{\rm{max}})$. For $^4$He, six datasets with
$\rm{max}(N_{\rm{max}})=10$ to $\rm{max}(N_{\rm{max}})=20$ are given,
providing the neural network with a sequence of mounting
information. The extrapolation result for the single neural network is
given by the prediction of $N_{\rm{max}}=100$ when the observable
value is virtually constant in the interval $10~\rm{MeV}< \hbar\omega
< 60~\rm{MeV}$. With each dataset, the multiple neural network
(containing 100 networks) is trained with randomly initialized network
values. The distribution of the multiple neural network results is
then fitted by the Gaussian function. Finally, the recommended values
of the multiple neural networks are set to be the mean value $\mu$ and
the uncertainties are defined as the standard deviation $\sigma$ of
the Gaussian.

Figure~\ref{fig:different_Nmax_observables_He4} shows the predictions
and corresponding uncertainties for the neural network approach
compared with the values obtained from the infrared (IR)
extrapolations of Ref.~\cite{forssen2018}. The error bars reflect the
variations that are due to changes in the initial point in the
training process. As we can see, the uncertainty of the neural network
predictions decreases with increasing $\rm{max}(N_{\rm{max}})$. This
indicates that the network is learning the pattern as the data set is
enlarged. The neural networks reach convergence after
$\rm{max}(N_{\rm{max}})=16$ and their predictions agree with the IR
extrapolations for both the ground-state energy and point-proton
radius.  We note that the two extrapolation methods exhibit different
behaviors while reaching identical converged values.

\begin{figure}[hbt]
  \includegraphics[width=0.48\textwidth]{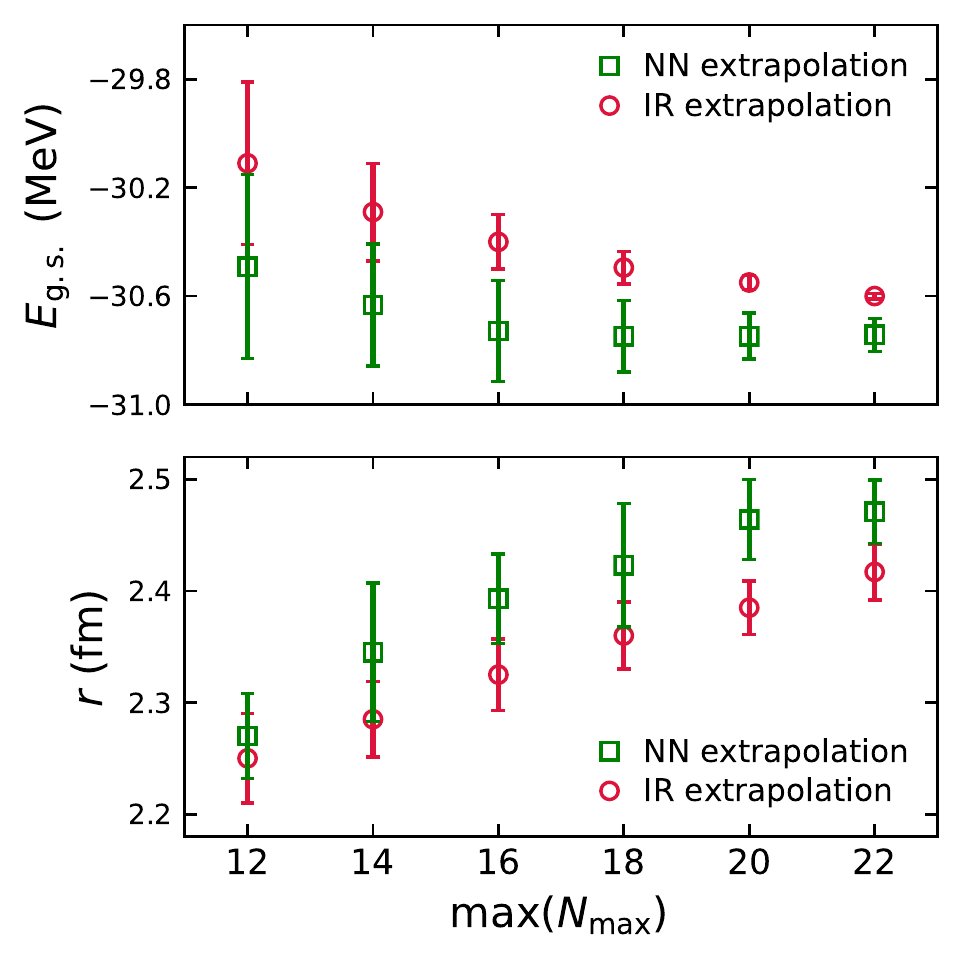}
  \caption{(Color online) Extrapolated results for $^6$Li ground-state
    energy (upper panel) and point-proton radius (lower panel) with
    NCSM datasets from $\rm{max}(N_{\rm{max}})=12$ to
    $\rm{max}(N_{\rm{max}})=22$ employing neural network (squares) and
    IR (circles) extrapolation. Error bars represent the uncertainties
    of the extrapolations that are due to changes in the initial point
    in the training process. }
  \label{fig:different_Nmax_observables_Li6}
\end{figure}

$^6$Li is a more challenging task for both $ab~initio$ calculations
and extrapolations. This is a weakly bound nucleus where a weakly
bound deuteron orbits the $^4$He core.  Thus, the radius is relatively
large, and the calculated observables converge slowly as the model
space increases. This nucleus is a good challenge for extrapolation
methods.  The results for neural network extrapolations are shown in
Figure~\ref{fig:different_Nmax_observables_Li6}. For the ground-state
energy, the neural network gives $E_{\rm{g.s.}}=-30.743\pm 0.061
~\rm{MeV}$ with the largest dataset $\rm{max}(N_{\rm{max}})=22$ and
the results start to converge when $\rm{max}(N_{\rm{max}})$ reaches
16. As a long-range operator the radius converges even slower than the
energy, which makes it more difficult for the extrapolation method to
obtain a reliable prediction. With the largest dataset, the neural
network extrapolated result is $r_{\rm{p}}=2.471\pm 0.028 ~\rm{fm}$
and the predictions start to converge at $\rm{max}(N_{\rm{max}})=20$.
The error bars reflect the variations that are due to changes in the
initial point in the training process.

So far, we have only studied the uncertainties from the random
starting point when training the network. To study the robustness of
the trained neural networks, we proceed as follows. Once a network is
trained, i.e. once its weights and biases $w$ are determined, we take
a random vector (with components drawn at random from a Gaussian
distribution with zero mean) $\Delta w$ in the space of weights and
biases and adjust its length such that the loss function fulfills
$L(w+\Delta w) = c L(w)$, with $c=2$ or $c=10$. These values are
motivated as follows. For a chi-square distribution with uncorrelated
degrees of freedom, $c=2$ would map out the region of one standard
deviation. However, our networks are not that simple and network
parameters are correlated. For this reason we also consider the case
$c=10$. We note that this approach yields uncertainty estimates but
not quantified uncertainties. We then use the new network parameters
$w+\Delta w$ to predict the observable of interest. Taking 100 random
vectors $\Delta w$ for each single network, we compute the variance in
the observable of interest, and also record the maximum deviation. The
results are shown in Tables~\ref{tab:tabnew1} and \ref{tab:tabnew2}
for $c=2$ and $c=10$, respectively. We see that the network is
approximately parabolic at its optimal training point (as variances
and maximal deviations increase by about a factor $\sqrt{5}$ as we go
from $c=2$ to $c=10$. For energies and radii, the networks are
robust. For $c=2$ and $c=10$, the network parameters $|\Delta w|/|w|$
change by about one per mill and one percent, respectively. Allowing
for a twofold increase of the loss function, the uncertainty from the
training of the network does not exceed the uncertainties from the
random initial starting points. However, allowing weights and biases
to change such that the loss function is increased by a factor of ten,
yields larger uncertainties. In this case, the maximum uncertainties
from the neural network (when added to the errorbars shown in
Fig.~\ref{fig:different_Nmax_observables_Li6}), would lead the
errorbars from the neural network extrapolation to overlap with those
from the IR extrapolation. We note finally that the single-layer
neural networks we employ are not resilient with regard to
dropout. Removing a single node after training of the network on
average changes the predictions for energies and radii by almost 20\%.

\begin{table}
\caption{
\label{tab:tabnew1}
Uncertainty analysis of NN extrapolated results for $^6$Li with
weights $w+\Delta w$. The random vector $\Delta w$ of weights and
biases is adjusted to double the loss function, i.e.  $L(w+\Delta w) =
2 L(w)$. The quantities $\sigma _{E_{\text{g.s.}}}$ and $\sigma _{r}$
are the standard deviation of the new predictions for ground-state
energy (in MeV) and point-proton radius (in fm),
respectively. $\text{max}(\Delta E_{\text{g.s.}})$ (in MeV) and
$\text{max}(\Delta r)$ (in fm) show the maximal deviation between the
new predictions and the origin results. $|\Delta w| / |w|$ are the
ratio between norms of the weights deviation and the origin weights.
}
\begin{ruledtabular}
\begin{tabular}{l cccccc p{cm}}
&&& \multicolumn{2}{c}{max($N_\text{{max}}$)}  &&\\
\cline{2-7}
 & 12 &14 & 16 & 18 &20&22\\
\colrule
 $\sigma _{E_{\text{g.s.}}}$       & 0.013    & 0.010   & 0.009  & 0.008  & 0.009  & 0.006  \\
 $\text{max}(\Delta E_{\text{g.s.}})$      & 0.068    & 0.049   & 0.037  & 0.031  & 0.032  & 0.024  \\
 $|\Delta w|  /  |w|$    & 0.0009   & 0.0008  & 0.0008 & 0.0008 & 0.0008 & 0.0008 \\
\hline
 $\sigma _{r}$     & 0.0034   & 0.0038  & 0.0042 & 0.0049   & 0.0054 & 0.0061 \\
 $\text{max}(\Delta r)$  & 0.0152   & 0.0176  & 0.0200 & 0.0212 & 0.0233 & 0.0269 \\
 $|\Delta w|  /  |w|$ & 0.0050   & 0.0043  & 0.0037 & 0.0042 & 0.0043   & 0.0030\\
\end{tabular}
\end{ruledtabular}
\end{table}

\begin{table}
\caption{
  \label{tab:tabnew2}
  Same as as Table~\ref{tab:tabnew1} but for random vectors $\Delta w$
  of weights and biases that yield a tenfold increase of the loss
  function.}
  \begin{ruledtabular}
\begin{tabular}{l cccccc p{cm}}
&&& \multicolumn{2}{c}{max($N_\text{{max}}$)}  &&\\
\cline{2-7}
 & 12 &14 & 16 & 18 &20&22\\
\colrule
 $\sigma _{E_{\text{g.s.}}}$ & 0.037    & 0.025   & 0.023  & 0.020  & 0.035  & 0.016  \\
$\text{max}(\Delta E_{\text{g.s.}})$  & 0.183    & 0.121   & 0.098  & 0.084  & 0.153  & 0.066  \\
$|\Delta w|  /  |w|$     & 0.0025   & 0.0023  & 0.0021 & 0.0021 & 0.0023 & 0.0021 \\
\hline
$\sigma _{r}$          & 0.0093   & 0.0093  & 0.0115 & 0.0139   & 0.0154   & 0.0165\\
$\text{max}(\Delta r)$ & 0.0415   & 0.0393  & 0.0525 & 0.0635 & 0.0684   & 0.0723\\
$|\Delta w|  /  |w|$ & 0.0122   & 0.0096  & 0.0105 & 0.0105 & 0.0114   & 0.0090\\
\end{tabular}
\end{ruledtabular}
\end{table}

\begin{figure}[t]
  \includegraphics[width=0.45\textwidth]{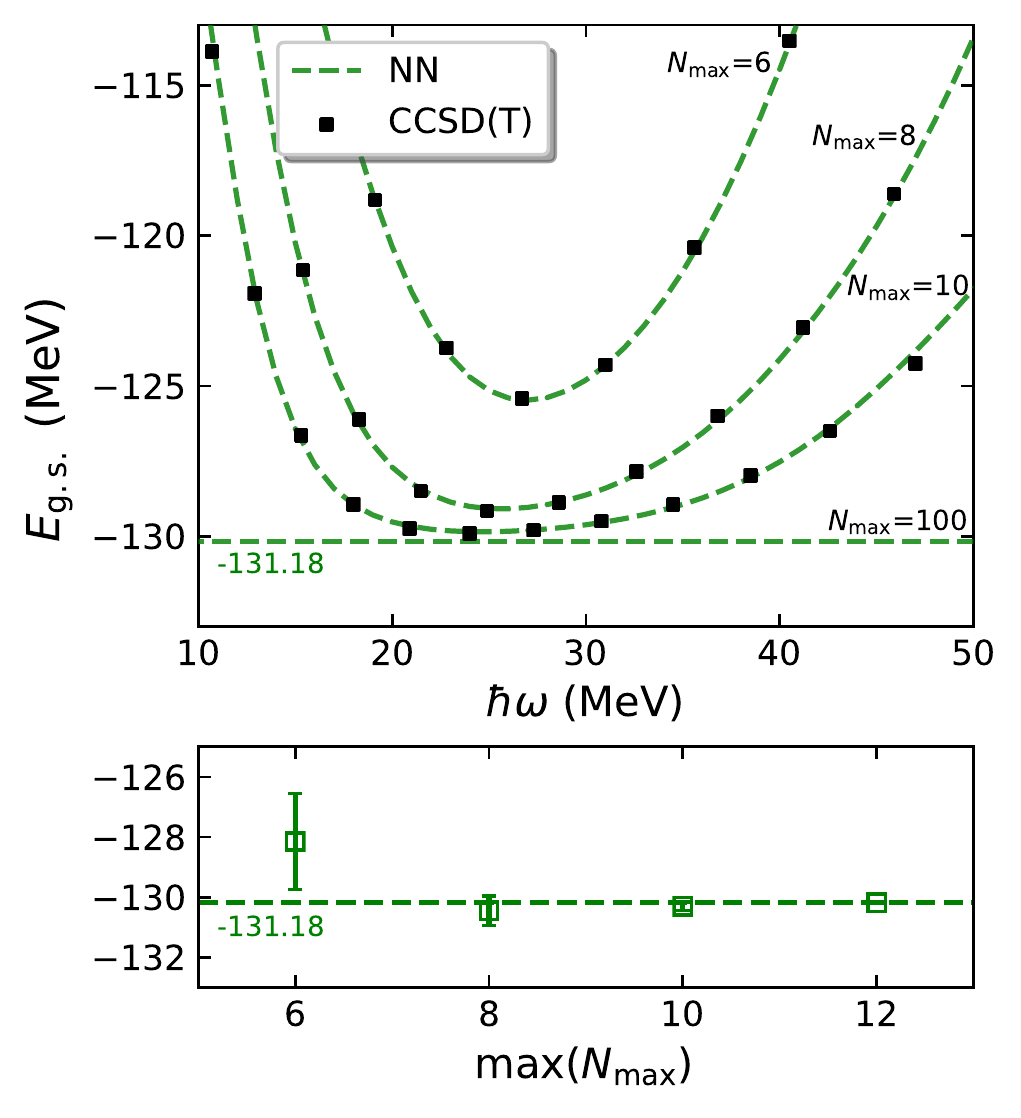}
  \caption{(Color online) neural network predictions (upper panel) based on the
    CCSD(T) dataset with $\rm{max}(N_{\rm{max}})=12$ and multiple neural network
    extrapolated results (lower panel) with datasets form
    $\rm{max}(N_{\rm{max}})=6$ to $\rm{max}(N_{\rm{max}})=12$.}
  \label{fig:different_Nmax_observables_O16}
\end{figure}

To illustrate the universality of neural network extrapolation, we
apply the multiple neural network approach on the ground-state energy
of $^{16}$O, computed with the coupled-cluster
method~\cite{forssen2018}. The upper panel of
Fig.~\ref{fig:different_Nmax_observables_O16} shows the neural network
performance with the largest datasets
[$\rm{max}(N_{\rm{max}})=12$]. As we can see in the lower panel of the
figure, the neural network extrapolation results start to converge at
$\rm{max}(N_{\rm{max}})=8$. Note that, by then, the neural network is
trained with only three sets of $N_{\rm{max}}$ data and still be able
to capture the correct pattern. This is due to the quick convergence
of the coupled-cluster method itself and the relatively flat curve
around the minimum of the energy as a function of $\hbar \omega$,
which are both favorable for the neural network extrapolation
approach.

\section{Summary}
In this paper, we presented a neural network extrapolation method to
estimate the ground-state energies and point-proton radii from NCSM
and the coupled-cluster calculations. To counter the overfitting
problem which is caused by the limited set of $ab~initio$ results, we
enlarged the data set by interpolating between different data points,
and used a loss function that accounts for the correlations between
the data points.  Because of the random nature of the neural network
algorithm, we employed multiple neural network approach to obtain
recommended results and uncertainties of the extrapolations. We
applied balanced sample weights as data preprocessing to eliminate the
influences of the persistent local minima, and to obtain a more
pronounced single solution for the multiple neural network
predictions.

We presented neural-network-extrapolated energies and radii of
$^{4}$He, $^{6}$Li for NCSM and compared them with IR extrapolated
results from Ref.~\cite{forssen2018}. The neural network
extrapolations gave reliable predictions for both observables with
reasonable uncertainties. The extrapolations for the ground-state
energy of $^{16}$O from coupled-cluster calculations also yielded
accurate results. The strong pattern learning ability of the neural
network allowed us to apply the same network architecture for NCSM and
CC extrapolation without employing any particular functions. In
conclusion, the neural networks studied in this work are useful tools
for extrapolating results from \emph{ab initio} calculations performed
in finite model-spaces.

\begin{acknowledgments}
  We thank Andreas Ekstr{\"o}m, Christian Forss{\'e}n, Dick Furnstahl
  and Stefan Wild for very useful and stimulating
  discussions. Weiguang Jiang acknowledges support as an FRIB-CSC
  Fellow. This material is based upon work supported in part by the
  U.S. Department of Energy, Office of Science, Office of Nuclear
  Physics, under Award Numbers DE-FG02-96ER40963 and DE-SC0018223. Oak
  Ridge National Laboratory is managed by UT-Battelle for the
  U.S. Department of Energy under Contract No. DE-AC05-00OR22725.
\end{acknowledgments}


%

\end{document}